\documentclass[aps,prb,twocolumn,floatfix,10pt, longbibliography]{revtex4-1}

\usepackage{multirow}

\newcommand{\bwt}{\begin{widetext}}
\newcommand{\ewt}{\end{widetext}}

\usepackage{float}
\usepackage{graphics,graphicx,amsmath}
\usepackage{tabularx,float}
\usepackage{epsfig}

\usepackage{bm,bbold}
\usepackage{color}
\usepackage{verbatim}

\usepackage[T1]{fontenc}
\usepackage{libertine}

\begin{document}

\title{Electronic Band Structure of Transition Metal Dichalcogenides from {\it ab initio} and Slater-Koster Tight-Binding Model}

\author{Jose \'Angel Silva-Guill\'en$^{1}$, Pablo San-Jose$^2$, and Rafael Rold\'an$^2$}
\affiliation{$^1$Fundaci\'on IMDEA Nanociencia, C/Faraday 9, Campus Cantoblanco, 28049 Madrid, Spain}
\affiliation{$^2$Instituto de Ciencia de Materiales de Madrid,
CSIC, Sor Juana Ines de la Cruz 3, 28049 Cantoblanco, Madrid, Spain}

\begin{abstract}

Semiconducting transition metal dichalcogenides present a complex electronic band structure with a rich orbital contribution to their valence and conduction bands. The possibility to consider the electronic states from a tight-binding model is highly useful for the calculation of many physical properties, for which first principle calculations are more demanding in computational terms when having a large number of atoms.  Here we present a set of Slater-Koster parameters for a tight-binding model that accurately reproduce the structure and the orbital character of the valence and conduction bands of single layer  {\em MX}$_2$, where $M=$ Mo, W and $X=$ S, Se. The fit of the analytical tight-binding Hamiltonian is done based on band structure from {\it ab initio} calculations. The model is used to calculate the optical conductivity of the different compounds from the Kubo formula.  

\end{abstract}

\date{\today}
\maketitle

\section{Introduction}

Soon after the discovery of graphene by mechanical exfoliation, this technique was applied to the isolation of other families of van der Waals materials.\cite{NG05} Among them, semiconducting transition metal dichalcogenides (TMD) are of special interest because they have a gap in the optical range of the energy spectrum, what makes them candidates for applications in photonics and optoelectronics.\cite{WS12,JH14,GZ14}  The electronic properties of these materials are highly sensitive to the external conditions such as strain, pressure or temperature. For instance, a direct-to-indirect gap and even a semiconducting-to-metal transition can be induced under specific conditions.\cite{FL12,PZ12,PV12,SS12,Molina-Sanchez_2015,Brumme2015}
 They also present a strong spin-orbit coupling (SOC) that, due to the absence of inversion symmetry in single layer samples, lifts the spin degeneracy of the energy bands.\cite{ZCS11}
In time reversal-symmetric situations, inequivalent valleys have opposite spin splitting,
leading to the so called spin-valley coupling,\cite{XY12,XH14,Touski_2016}
which has been observed experimentally.\cite{ZC12,MH12,CF12,ZC13,WX13,WS13} The coupling of the spin and valley degrees of freedom in semiconducting TMDs opens the possibility to manipulate them  for future applications in spintronics and valleytronics nano-devices.\cite{MS13,SU12,ZC12,KB13,Kormanyos2015}

On the other hand, TMDs present a high stretchability. Moreover, external strain can be used to efficiently manipulate their electronic and optical properties.\cite{RG15} Non-uniform strain profiles can be used to create {\it funnels of excitons}, that allows to capture a broad light spectrum, concentrating carriers in specific regions of the samples.\cite{FL12,San-Jose_2016}  Strain engineering can be also used to exploit the piezoelectric properties of atomically thin layers of TMDs, converting mechanical to electrical energy. \cite{WW14}

The rich orbital structure of the valence and conduction bands of semiconducting TMDs\cite{Roldan_AP_2014} complicates the construction of a tight-binding (TB) model for these systems. Such a TB model must be precise enough as to include all the pertinent orbitals of the relevant bands, but at the same time, simple enough as to be used without too much effort in calculations of optical and transport properties of these materials. The advantage of a tight-binding description with respect to first-principles methods is that it provides a simple starting point for the further inclusion of many-body electron-electron interaction, external strains, as well as of the dynamical effects of the electron-lattice interaction. Tight-binding approaches are often more convenient than {\it ab initio} methods for investigating systems involving a very large number of atoms,\cite{San-Jose_2016} disordered and inhomogeneous samples,\cite{YG14} strained and/or bent samples,\cite{Rostami2015,Pearce2015} materials nanostructured in large scales (nanoribbons,\cite{RAG15,Farmanbar2016} ripples\cite{CS13}) or in twisted multilayer materials. The aim of the present paper is twofold. Starting from the TB model for MoS$_2$ developed by Cappelluti {\it et al.},\cite{CG13} we present a more accurate set of Slater-Koster parameters obtained from a more sophisticated fitting procedure, and we further generalize it to the other families of semiconducting TMDs, namely WS$_2$, MoSe$_2$ and WSe$_2$. Finally, we apply the obtained tight-binding models to calculate the optical conductivity of the four compounds.

\section{Electronic Band Structure}

\begin{figure}[t]
\includegraphics[scale=1.,clip=]{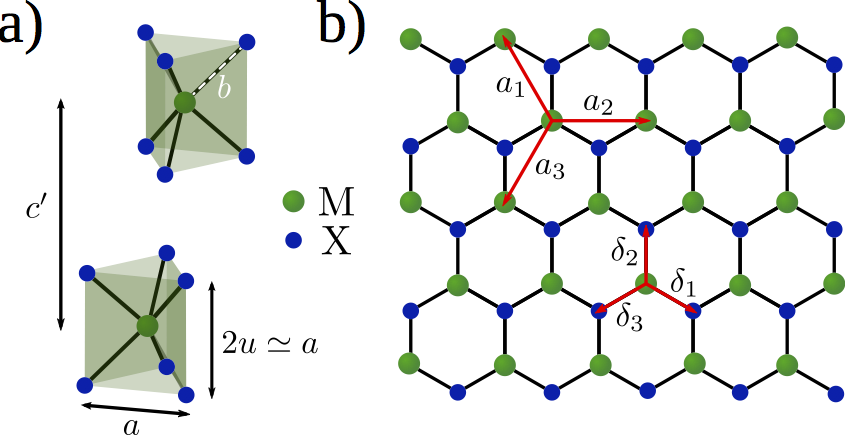}
\caption{a) Sketch of the atomic structure of {\em MX}$_2$.
The bulk compound has a 2H-{\em MX}$_2$ structure
with two {\em MX}$_2$  layers per unit cell, each of them
being built up from a trigonal prism coordination unit.
The value of the lattice constants for each family is given in Table \ref{Tab:LattParam}. b) Top view of monolayer $MX_2$ lattice. Green (Blue) circles indicate $M$ ($X$) atoms. The nearest neighbors ($\delta_i$) and the next nearest neighbors ($a_i$) vector are shown in the figure.
}
\label{Fig:Structure}
\end{figure}

\begin{figure}[t]
\begin{center}
\includegraphics[width=0.4\linewidth]{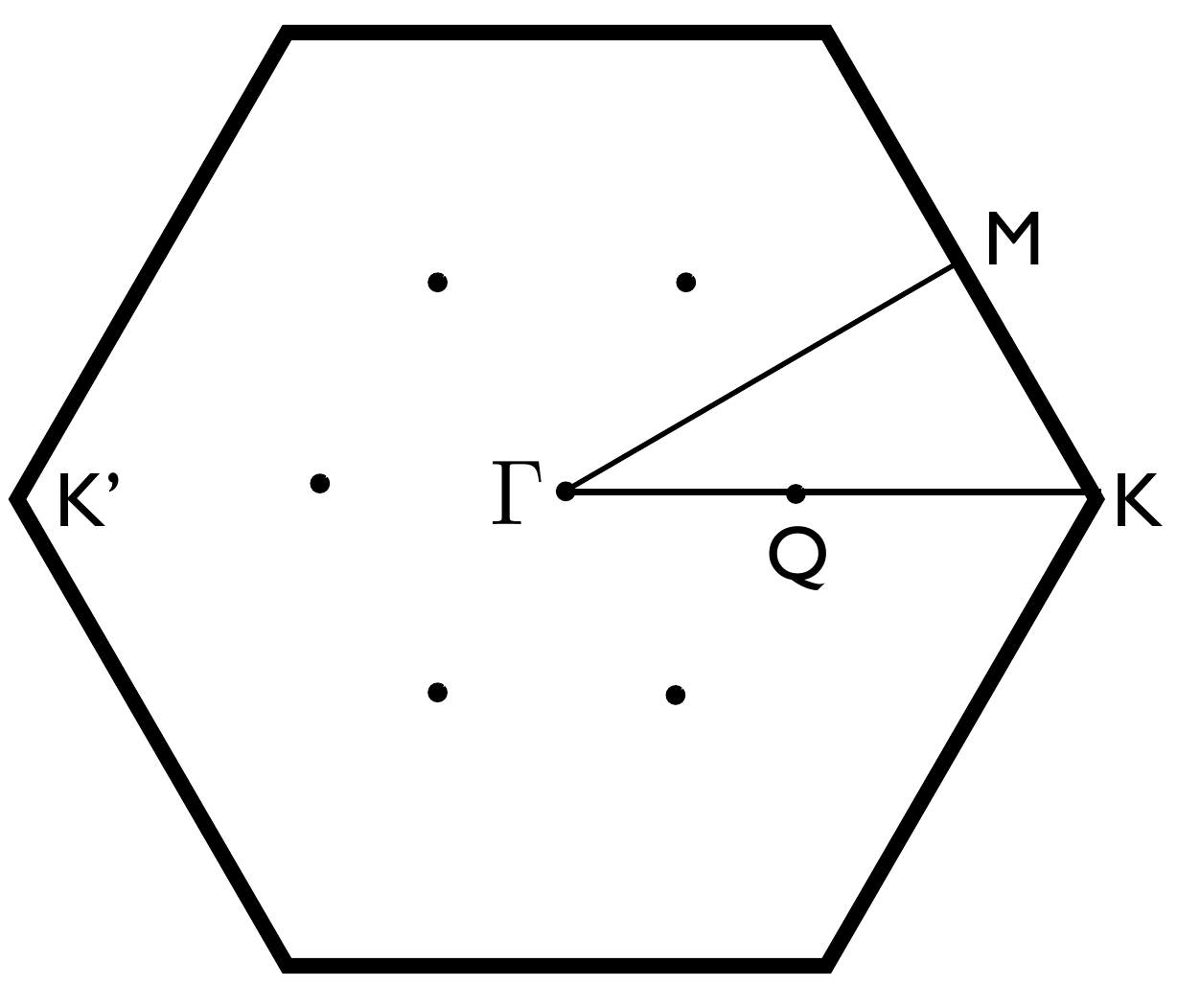}
\end{center}
\caption{Two dimensional Brillouin zone of $MX_2$. The high symmetry points $\Gamma=(0,0)$,  K$=4\pi/3a(1,0)$ and M$=4\pi/3a (0,\sqrt{3}/2)$ are shown. The Q points (which are not high symmetry points) indicate the position of the edges of the conduction band in multi-layer samples.} \label{Fig:BZ}
\end{figure}

The crystal structure of  {\em MX}$_2$ 
is schematically shown in Fig. \ref{Fig:Structure}. A single layer is composed by an inner layer of metal $M$ atoms ordered
on a triangular lattice, which is sandwiched between two layers
of chalcogen $X$ atoms placed on the triangular lattice of alternating
hollow sites. We use a notation such that $a$ corresponds to the distance between
nearest neighbor in-plane $M-M$ and $X-X$ atoms,
$b$ is the nearest neighbor $M-X$ separation and
$u$ is the distance between the $M$ and $X$ planes.
The {\em MX}$_2$ crystal forms an almost perfect trigonal prism
structure with $b \simeq \sqrt{7/12}a$ and $u \simeq a/2$. The lattice parameters of the bulk
compounds corresponding to the more commonly studied TMDs are given in Table \ref{Tab:LattParam}.\cite{LX13,KGF13,KA12} The in-plane Brillouin zone is an hexagon, and it is shown in Fig. \ref{Fig:BZ}. It contains the
high-symmetry points $\Gamma=(0,0)$, K$=4\pi/3a(1,0)$ and M$=4\pi/3a (0,\sqrt{3}/2)$. The six Q points correspond to the approximate position of a local minimum of the conduction band.

\begin{table}[b]
\begin{tabular}{|l||c|c|c|}
\hline
\hline
 &  $a$ & $u$ & $c'$ \\
\hline
MoS$_2$ & $3.160$ & $1.586$ & $6.140$ \\
\hline
MoSe$_2$ & $3.288$ & $1.664$ & $6.451$ \\
\hline
WS$_2$ & $3.153$ & $1.571$ & $6.160$ \\
\hline
WSe$_2$ & $3.260$ & $1.657$ & $6.422$ \\
\hline
\hline
\end{tabular}
\caption{
Lattice parameters for the TMDs considered in the text. $a$ represents the $M$-$M$ atomic distance,
$u$ the internal vertical separation between the $M$
and the $X$ planes, and $c'$ the distance between the metal $M$ layers.
In bulk systems $c=2c'$ accounts for the $z$-axis lattice parameter.
All values are in \AA\, units.
}
\label{Tab:LattParam}
\end{table}

\begin{figure}[t]
\includegraphics[scale=0.3,clip=]{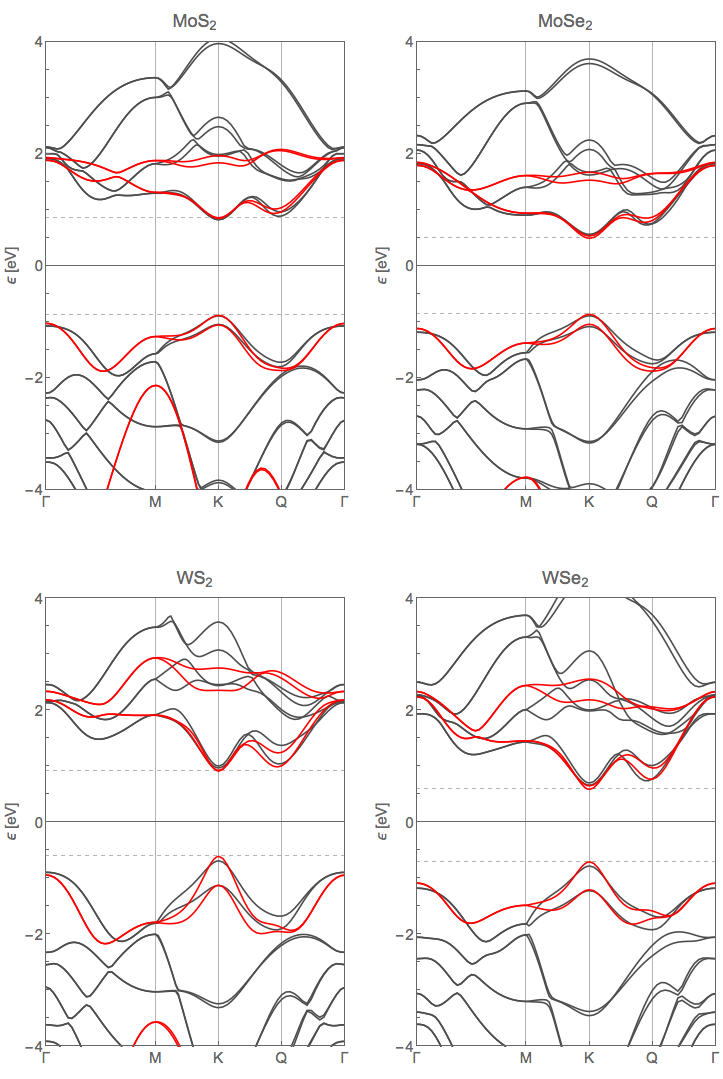}
\caption{Electronic band structure of single-layer $MX_2$ from DFT calculations (black lines) and from tight-binding (red lines).
}
\label{Fig:Fitting}
\end{figure}

In order to study the electronic band structure of single-layer TMDs, we use the Density Functional Theory (DFT) calculations presented by some of the authors in Ref. \onlinecite{Roldan_AP_2014}, in which the intrinsic spin-orbit interaction term for all atoms is included.  
Fig. \ref{Fig:Fitting} shows the band structures for single-layer $MX_2$ (black lines) together with the TB bands that will be discussed later (red lines).\footnote{DFT calculations were done using
the \textsc{Siesta} code,\cite{SS02,AS08} with the exchange-correlation potential of Ceperly-Alder\cite{CA80}
as parametrized by Perdew and Zunger.\cite{PZ81}
A split-valence double-$\zeta $ basis set including
polarization functions was considered.\cite{AS99} 
The energy cutoff and the BZ sampling were chosen to converge the total energy with a value of 300 Ry and $30\times30\times1$. The energy cutoff was set to $10^{-6}$ eV.
Spin-orbit interaction of the different compounds was considered by following the method developed in Ref. \onlinecite{FF06}. 
The lattice parameters used in the calculation are given 
in Table \ref{Tab:LattParam}.}
One of the main characteristics of TMDs is that, contrary to what happens in other 2D crystals like graphene or phosphorene, the valence and conduction bands of $MX_2$ present a very rich orbital contribution. As explained in detail in Ref. \onlinecite{CG13}, they are made by hybridization of the $d$ orbitals of the transition metal, and the $p$ orbitals of the chalcogen. More specifically, the analysis of the orbital content of the set of bands containing the first four conduction bands and the first seven valence bands, which cover an energy window from -7 to 5 eV around the Fermi level, approximately, reveals that these bands are dominated by the five 4$d$ (5$d$) orbitals of the metal Mo (W) and
the six (three for each layer) 3$p$ (4$p$) orbitals of the chalcogen S (Se), 
summing up to the $93$ \% of the total orbital weight.\cite{CG13}

\begin{figure*}[t]
\includegraphics[scale=0.28,clip=]{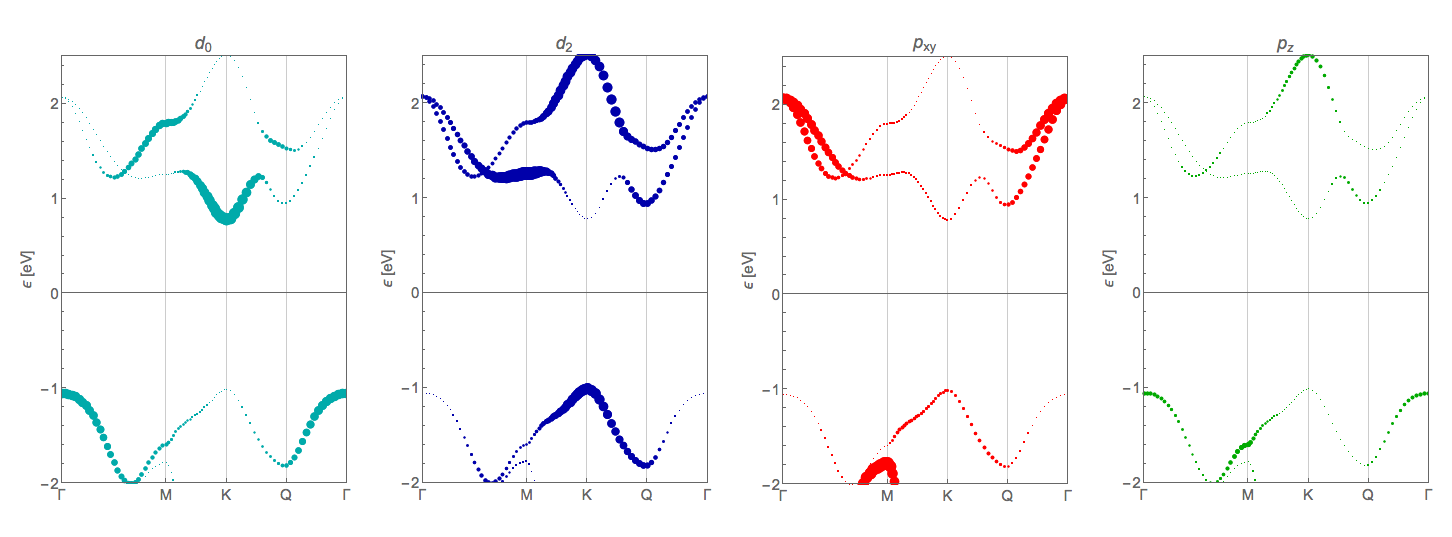}
\includegraphics[scale=0.28,clip=]{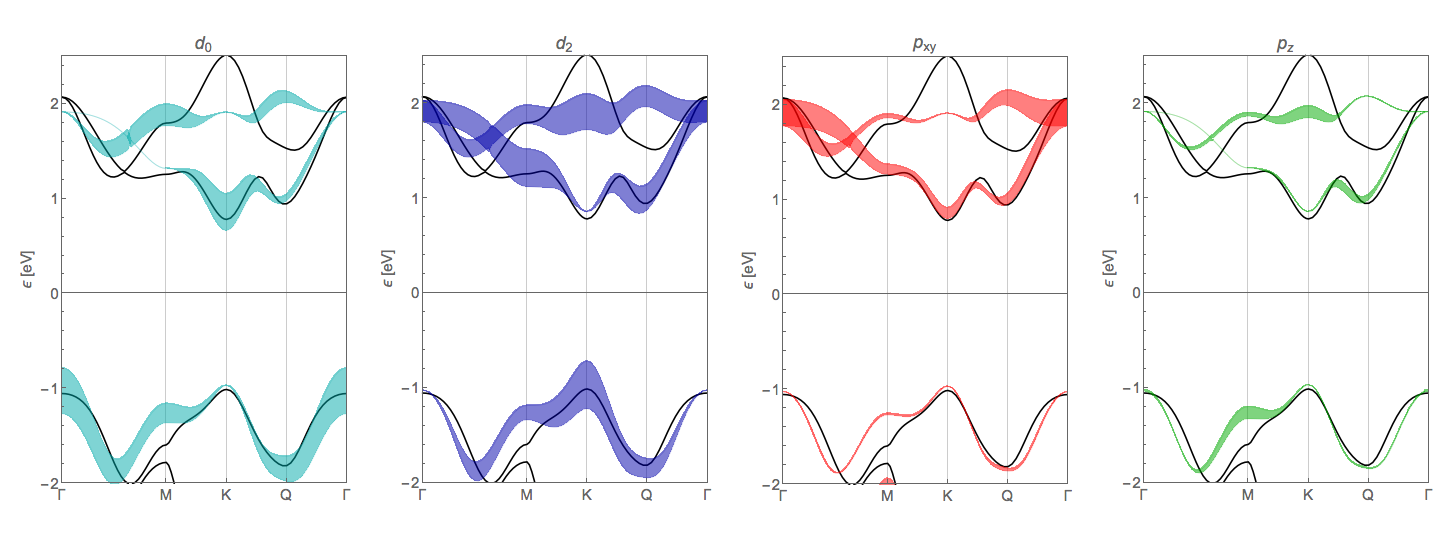}
\caption{Band structure and orbital character of single-layer MoS$_2$. The thickness of the bands represents the orbital weight, where the $d$ character ($d_2 =d_{x^2-y^2},d_{xy}$ and $d_0 = d_{3z^2-r^2}$) refers to the Mo atom $4d$ orbitals, while the $p$ character (where $p_{xy} =p_x,p_y$) refers to $3p$ orbitals of sulfur. Top panels correspond to orbital weight from DFT calculations, whereas bottom panels correspond to orbital weight from TB results. The black lines in the bottom panels are the DFT bands. Notice that spin-orbit coupling is not included in these plots.
}
\label{Fig:FB-MoS2}
\end{figure*}

Single-layer TMDs are direct gap semiconductors, with the gap located at the two inequivalent K and K' points of the Brillouin zone (Fig. \ref{Fig:Fitting}). The main orbital character at the edge of the valence band is due to a combination of  $d_{x^2-y^2}$ and $d_{xy}$ orbitals of the metal $M$, which hybridize with $p_x$ and $p_y$ orbitals of the chalcogen $X$. On the other hand, the edge of the conduction band is formed by $d_{3z^2-r^2}$ orbital of $M$, plus some contribution of $p_x$ and $p_y$ orbitals of $X$.\cite{CG13} Contrary to single-layer samples, multi-layer compounds are indirect gap semiconductors. The edge of the valence band lies at the $\Gamma$ point of the BZ, having a major contribution from $d_{3z^2-r^2}$ and $p_z$ orbitals of $M$ and $X$ atoms, respectively. The edge of the conduction band in multi-layer samples is placed at the Q point of the BZ. It is important to notice that the Q point is not a high symmetry point of the Brillouin zone, and therefore its exact position depends on the number of layers and on the specific compound. The orbital character at the Q point originates mainly from the $d_{xy}$ and $d_{x^2-y^2}$ orbitals of the metal $M$, plus $p_x$ and $p_y$ orbitals of the chalcogen $X$, plus a non negligible contribution of $p_z$ and $d_{3z^2-r^2}$ of $X$ and $M$ atoms, respectively. Figs. \ref{Fig:FB-MoS2}-\ref{Fig:FB-WSe2} represent these relative orbital weights in detail for the different compounds. The extremely rich orbital contribution to the relevant bands that occur in semiconducting TMDs complicates the derivation of a minimal TB model, valid in the whole Brillouin zone. Another important feature of TMDs is that they present a strong SOC, that leads to a large splitting of the valence band at the K and K' points of the BZ (see Fig. \ref{Fig:Fitting}). The splitting is bigger for W than for Mo compounds, due to the heavier mass of the former. SOC also leads to a splitting of the conduction band at the K point, as well as at the minimum at  the Q point.\cite{KGF13,Roldan_2DMat_2014}

\section{Tight-binding model}

In this section we consider the electronic band structure of TMDs, in the whole BZ, from a Slater-Koster tight-binding approximation.\cite{SK54} We use the model developed by Cappelluti {\it et al.},\cite{CG13} which contains 11 orbitals per layer.
In particular, the model contains the five $d$ orbitals of the metal $M$ atom
and  the six $p$ orbitals of the two chalcogen $X$ atoms in the unit cell. The used scheme has been recently used in other works studying the electronic band structure of TMDs from a tight-binding perspective.\cite{Ridolfi2015,Fang2015} The corresponding base can be expressed as\cite{CG13}
\begin{equation}
\left(p_x^t,p_y^t,p_z^t,d_{3z^2-r^2},d_{xz},d_{yz},~d_{x^2-y^2},~d_{xy},p_{x}^b,~p_{y}^b,~p_{z}^b\right)
\end{equation}
where the indices $t$ and $b$ label the top and bottom chalcogen planes, respectively.
The model is defined by the hopping integrals between the different orbitals, which are
described in terms of $\sigma$, $\pi$ and
$\delta$ ligands. In the following we reproduce the most important results, and we refer the reader to Refs. \onlinecite{CG13,Roldan_2DMat_2014} for details of the model. The Slater-Koster parameters that account for the relevant hopping processes of the model are
$V_{pd\sigma}$ and $V_{pd\pi}$ for $M-X$ bonds,
$V_{dd\sigma}$, $V_{dd\pi}$ and $V_{dd\delta}$ for $M-M$ bonds,
and $V_{pp\sigma}$ and $V_{pp\pi}$ for $X-X$ bonds.
Additional parameters of the theory are the crystal fields
$\Delta_0$, $\Delta_1$,  $\Delta_2$, $\Delta_p$, $\Delta_z$, 
describing respectively the atomic level of the $l=0$ ($d_{3z^2-r^2}$), the $l=1$ ($d_{xz}$, $d_{yz}$), the $l=2$ ($d_{x^2-y^2}$, $d_{xy}$) $M$ orbitals, the in-plane ($p_x$, $p_y$) $X$ orbitals and of the out-of-plane $p_z$ $X$ orbitals.

This model can be simplified by performing an unitary transformation that takes the $p$ orbitals of the top and bottom layers of the $X$ atoms into their symmetric and antisymmetric combinations with respect to the $z$-axis. This way the 11-bands model is decoupled into a $6\times 6$ block with even (odd) symmetry of the $p_x,p_y$ ($p_z$) orbitals with respect to $z \rightarrow -z$ inversion, and a $5\times 5$ bands block with opposite combination.
Since low energy excitations belong exclusively to the first block, the fit to DFT that we will present later will be performed within this sector. Therefore the relevant bands above and below the gap are well accounted by 
the reduced Hilbert space:
\begin{equation}
\psi
=
\left(d_{3z^2-r^2},~d_{x^2-y^2},~d_{xy},p_{x}^S,~p_{y}^S,~p_{z}^A\right)
\label{Eq:BaseE}
\end{equation}
where the \emph{S} and \emph{A} superscripts  stand for the symmetric and antisymmetric combinations of the \emph{p}-orbitals of the $X$ atom, $p_i^{S}=1/\sqrt{2}(p_i^t+p_i^b)$ and $p_i^A=1/\sqrt{2}(p_i^t-p_i^b)$, with $i=x,y,z$. The tight-binding Hamiltonian defined by the base (\ref{Eq:BaseE}),
including local spin-orbit coupling,
can be expressed in real space as
\begin{eqnarray}
\label{Eq:H-Real}
H
&=&
\sum_{i,\mu\nu}
\epsilon_{\mu,\nu} c^\dagger_{i,\mu}c_{i,\nu}
+
\sum_{ ij,\mu\nu}
{[t_{ij,\mu\nu} c^\dagger_{i,\mu}c_{j,\nu}+{\rm H.c.}]},
\end{eqnarray} 
where $c^\dagger_{i,\mu}(c_{i,\mu})$ creates (annihilates) an electron in the unit cell $i$
in the atomic orbital $\mu=1,\ldots,6$, belonging to the Hilbert space
defined by (\ref{Eq:BaseE}). The Hamiltonian in $k$-space can be expressed as: \cite{CG13,Roldan_2DMat_2014,Rostami2015}
\begin{eqnarray}\label{Eq:H-k}
{\cal H}&=&\begin{pmatrix}{\cal H}_{MM}&&{\cal H}_{MX}\\{{\cal H}_{MX}}^\dagger&&{\cal H}_{XX}\end{pmatrix},\nonumber\\
{\cal H}_{MM}&=&\epsilon_{M}+2\sum_{i=1,2,3}{t^{MM}_i\cos \left({\bf k}\cdot{\bf a}_i\right)},\nonumber\\
{\cal H}_{XX}&=&\epsilon_{X}+2\sum_{i=1,2,3}{t^{XX}_i\cos\left( {\bf k}\cdot{\bf a}_i\right)},\nonumber\\
{\cal H}_{MX}&=&\sum_{i=1,2,3}{t^{MX}_i e^{-i {\bf k}\cdot{\bm \delta}_i}},
\end{eqnarray}
where the vectors (${\bm \delta}_i$) and (${\bf a}_i$) are shown in Fig. \ref{Fig:Structure}(c). The analytical expressions for the TB model are given in Appendix \ref{App:TB}.

\begin{table}[t]
\begin{tabular}{lclcrcrc}
\hline
\hline
                          &                                &                   &                                 &   \hspace{0.1truecm}  MoS$_2$      &\hspace{0.1truecm}  MoSe$_2$ &\hspace{0.1truecm}WS$_2$ &\hspace{0.1truecm} WSe$_2$ \\
\hline
\\
  SOC   & \hspace{0.1truecm} &$\lambda_M$ & \hspace{0.1truecm} & \hspace{0.1truecm} 0.086 &  \hspace{0.1truecm} 0.089 &\hspace{0.1truecm} 0.271 &\hspace{0.1truecm} 0.251   \\
             & \hspace{0.1truecm} &$\lambda_X$ & \hspace{0.1truecm}  &\hspace{0.1truecm} 0.052 & \hspace{0.1truecm}  0.256 & \hspace{0.1truecm} 0.057 & \hspace{0.1truecm}0.439  \\
\\
   Crystal Fields & \hspace{0.1truecm} &$\Delta_0$  & \hspace{0.1truecm} &  -1.094 & \hspace{0.1truecm}  -1.144 & -1.155 & -0.935 \\
                          & \hspace{0.1truecm} &$\Delta_1$ & \hspace{0.1truecm} &  -0.050         & \hspace{0.1truecm} -0.250   & -0.650 & -1.250\\
                          & \hspace{0.1truecm} &$\Delta_2 $ & \hspace{0.1truecm} & -1.511  & \hspace{0.1truecm}  -1.488  & -2.279 & -2.321\\
                          & \hspace{0.1truecm} &$\Delta_p$  & \hspace{0.1truecm} &  -3.559 & \hspace{0.1truecm}  -4.931 & -3.864& -5.629\\
                          & \hspace{0.1truecm} &$\Delta_z$ & \hspace{0.1truecm} &  -6.886 & \hspace{0.1truecm}  -7.503  & -7.327& -6.759\\
\\                
$M$-$X$ & \hspace{0.1truecm} &$V_{pd\sigma}$ & \hspace{0.1truecm} &  3.689 & \hspace{0.1truecm}  3.728 & 7.911 & 5.803 \\
                          & \hspace{0.1truecm} &$V_{pd\pi}$ & \hspace{0.1truecm} &  -1.241  & \hspace{0.1truecm}  -1.222 & -1.220 & -1.081\\
\\                      
$M$-$M$& \hspace{0.1truecm} &$V_{dd\sigma}$ & \hspace{0.1truecm} &  -0.895 & \hspace{0.1truecm}  -0.823  & -1.328& -1.129\\
                          & \hspace{0.1truecm} &$V_{dd\pi}$ & \hspace{0.1truecm} &  0.252 & \hspace{0.1truecm}   0.215  & 0.121& 0.094 \\
                          & \hspace{0.1truecm} &$V_{dd\delta}$ & \hspace{0.1truecm} &  0.228 & \hspace{0.1truecm}  0.192   & 0.442& 0.317\\
\\                        
$X$-$X$  & \hspace{0.1truecm} &$V_{pp\sigma}$ & \hspace{0.1truecm} &  1.225 & \hspace{0.1truecm}  1.256   & 1.178& 1.530\\
                          & \hspace{0.1truecm} &$V_{pp\pi}$ & \hspace{0.1truecm} &  -0.467 & \hspace{0.1truecm}  -0.205  & -0.273& -0.123\\                     
\hline
\hline
\end{tabular}
\caption{Spin-orbit coupling $\lambda_{\alpha}$ and tight-binding parameters for single-layer $MX_2$, where the metal $M$ is Mo or W and $X$ is S or Se. All the Slater-Koster parameters are in units of eV. SO coupling parameters are taken from Ref. \onlinecite{KGF13}.}
\label{Tab:Parameters}
\end{table}

\begin{table}[t]
\begin{tabular}{|cc|cc|cc|cc|}
\hline
\hline
& & \multicolumn{2}{c|}{K$^v$}& \multicolumn{2}{c|}{K$^c$}& \multicolumn{2}{c|}{$\Gamma^v$}  \\
\hline
& & DFT& TB & DFT & TB & DFT & TB \\
\hline
\multirow{4}{*}{MoS$_2$} & $d_0$ & 0.0 & 0.0  & 0.82 & 0.77 & 0.66 & 0.96 \\
 & $d_2$ & 0.76 & 1.0 & 0.0 & 0.0 & 0.0 & 0.0 \\
 & $p_{xy}$ & 0.20 & 0.0 & 0.12 & 0.23 & 0.0 & 0.0 \\
 & $p_z$ & 0.0 & 0.0 & 0.0 & 0.00 & 0.28 & 0.04 \\
\hline
\multirow{4}{*}{MoSe$_2$} & $d_0$ & 0.0 & 0.0 & 0.83 & 0.83 & 0.71 & 0.96 \\
 & $d_2$ & 0.78 & 1.0  & 0.0 & 0.0 & 0.0 & 0.0 \\
 & $p_{xy}$ & 0.18 & 0.0 & 0.10 & 0.17 & 0.0 & 0.0 \\
 & $p_z$ & 0.0 & 0.0 & 0.0 & 0.0 & 0.23 & 0.04 \\
 \hline
\multirow{4}{*}{WS$_2$}  & $d_0$ & 0.0 & 0.0 & 0.80 & 0.76 & 0.64 & 0.98 \\
& $d_2$ & 0.74 & 0.94 & 0.0 & 0.0 & 0.0 & 0.0 \\
& $p_{xy}$ & 0.21 & 0.06 & 0.07 & 0.24 & 0.0 & 0.0 \\
& $p_z$ & 0.0 & 0.0 & 0.0 & 0.0 & 0.28 & 0.02 \\
\hline
\multirow{4}{*}{WSe$_2$}  & $d_0$ & 0.0 & 0.0 & 0.82 & 0.86 & 0.69 & 0.99 \\
& $d_2$ & 0.73 & 0.95 & 0.0 & 0.0 & 0.0 & 0.0 \\
& $p_{xy}$ & 0.20 & 0.05 & 0.05 & 0.14 & 0.0 & 0.0 \\
& $p_z$  & 0.0 & 0.0 & 0.0 & 0.0 & 0.23 & 0.01 \\
\hline
\hline
\end{tabular}
\caption{Comparison of the orbital contribution at band edges at K  and $\Gamma$ points obtained from DFT and TB models.  K$^v$ (K$^c$) refers to the edge of the valence (conduction) band at K point, and $\Gamma^v$ refers to the edge of the valence band at the $\Gamma$ point.}
\label{Tab:Orbital}
\end{table}

\section{Slater-Koster parameters from fitting to {\it ab initio} calculations}

Finding the optimal set of Slater-Koster parameters for the TB model considered here is a difficult task. Two main requirements must be satisfied: a good reproduction of the structure of the relevant electronic bands, and faithful representation of the orbital contribution along such bands. The last condition is especially relevant because, for example, different kinds of strain do not affect all the hoppings in the same manner. Therefore capturing the proper orbital contribution is essential when using the TB model for calculations of physical properties of strained membranes. The same happens when one consider the effect of vacancies, adatoms, etc. 
 
In this work we have obtained the Slater-Koster parameters for each compound from a minimization procedure which has the possibility to consider a band/momentum resolved weight, that allows us to resolve more accurately particular $\bf k$ regions of selected bands (e.g. edges of the valence and conduction bands).\footnote{The calculation of tight-binding band structure and fitting to DFT have been performed using the MathQ package, developed by P. San-Jose. (Source: http://www.icmm.csic.es/sanjose/MathQ/MathQ.html).} Furthermore, we can apply constrictions for the orbital contribution at specific band regions, taking as a reference the information from the DFT wave-functions. The sets of Slater-Koster parameters that we have obtained for the four compounds are given in Table \ref{Tab:Parameters}. During the fitting we have used only the $6\times6$ block of the Hamiltonian because, as explained above, it accounts for the valence and conduction bands. Therefore, we obtain as output all the Slater-Koster parameters but one, $\Delta_1$, which is the crystal field corresponding to $d_2=d_{xz,yz}$ orbitals, whose contribution is absent in the $6\times 6$ block. What we have done to estimate the value of $\Delta_1$ is to impose that the edges of the TB and DFT bands coincide for the lower energy band of the $5\times 5$ block  at the K point of the Brillouin zone.  
The band structure calculated with the $6\times 6$ block of this model is plotted in Fig. \ref{Fig:Fitting}, as compared to DFT calculations. In Figs. \ref{Fig:FB-MoS2},\ref{Fig:FB-MoSe2}-\ref{Fig:FB-WSe2} we compare the orbital contribution of the TB model, using the Slater-Koster parameters of Table \ref{Tab:Parameters}, to the corresponding orbital contribution as obtained from DFT. We show the results for the most relevant orbitals ($d_0=d_{3z^2-r^2}$, $d_2=d_{xy},d_{x^2-y^2}$, $p_{xy}=p_x,p_y$ and $p_z$), and we can conclude that the TB model not only present an acceptable fit to the band structure, but importantly, the wave-functions also reproduce the DFT orbital contribution at the most important points of the band structure. Table \ref{Tab:Orbital} contains the main orbital contribution of each compound at the most relevant edges of the band structure, namely valence and conduction bands at K point, and valence band at $\Gamma$ point of the Brillouin zone.  We notice that the main restriction of the TB model considered here is that it only includes up to next-nearest-neighbor hopping terms, and this is why the fit to the DFT bands cannot be perfect. More sophisticated methods as DFT based tight-binding Hamiltonians represented in the basis of maximally localized Wannier functions can lead to better agreements, at the cost of inclusion of longer range hopping terms.\cite{Lado_2016} For the case presented here, and due to the automatised fitting procedure, we can conclude that the set of parameters presented in Table \ref{Tab:Parameters} must be close to the ideal solution.

\begin{figure}[t]
\includegraphics[scale=0.9,clip=]{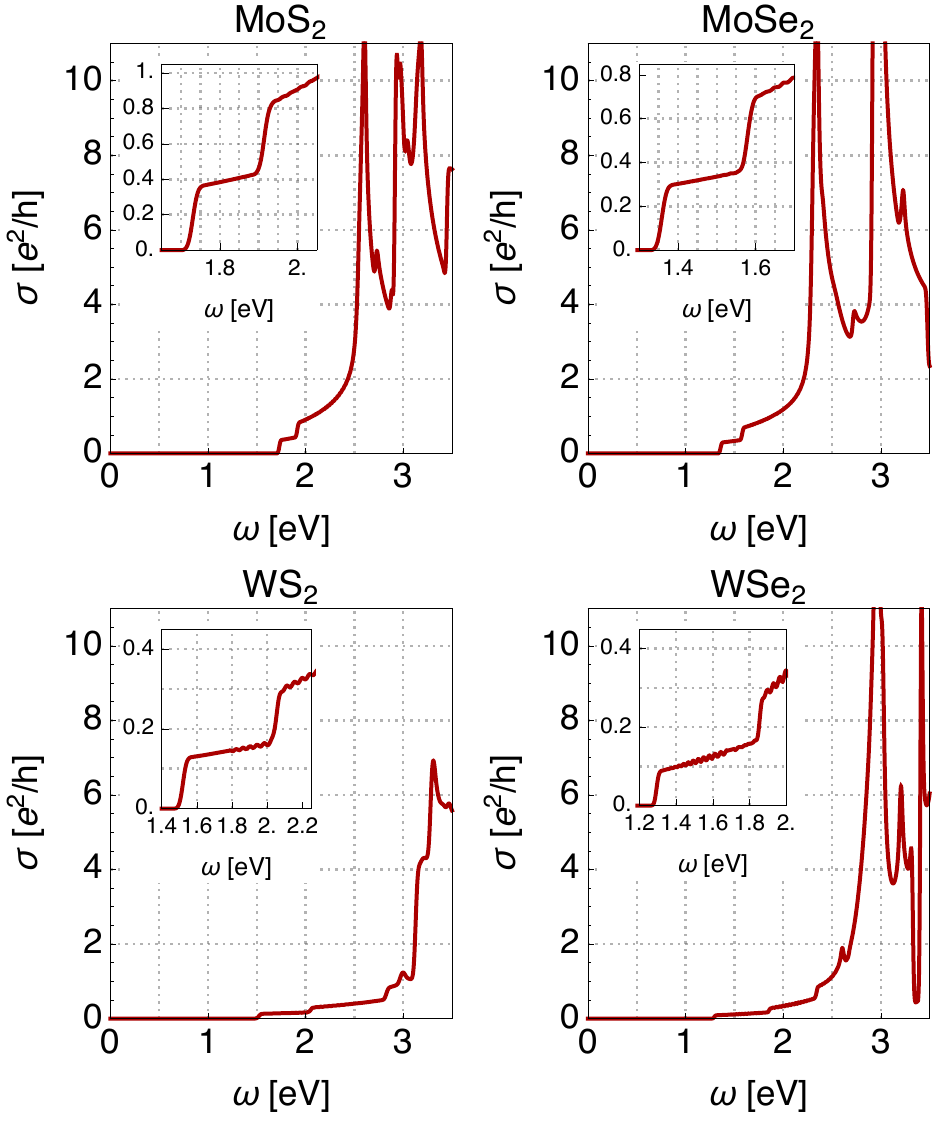}
\caption{Optical conductivity of the four compounds, calculated from the TB model. The insets show the respective low energy zooms of $\sigma(\omega)$ around the onset of optical transitions. 
}
\label{Fig:OpticalConductivity}
\end{figure}

\section{Optical Conductivity}\label{Sec:OptCond}

Once we have the tight-binding models for the four compounds, we can use them to calculate physical observables as, for example, the optical conductivity $\sigma(\omega)$. For this aim we use the Kubo formula 
\begin{eqnarray}
\sigma(\omega)&=&\frac{e^2}{h}\frac{\hbar}{A\omega}\sum_{mn,{\bf k}}(f(E_n)-f(E_m))\times\nonumber\\
&&\left|  \left \langle \Psi_n({\bf k})| \hat v | \Psi_m({\bf k}) \right \rangle \right |^2 
\delta [\hbar \omega - (E_n({\bf k})-E_m({\bf k}))]\nonumber\\
\end{eqnarray}
where $A$ is the area of the unit cell, $\Psi_n({\bf k})$ is the eigenstate of energy $E_n$, $f(E_n)=1/(1+e^{\beta E_n})$ is the Fermi-Dirac distribution function, in terms of the inverse temperature $\beta=1/k_BT$ and considering that the Fermi energy lies in the gap, and $\hat v = (1/\hbar) \partial \hat {\cal H}/\partial {\bf k}$ is the velocity operator. The results are shown, for the approximate range of validity of our TB models ($\sim 1$~eV above and below the band gap), in Fig. \ref{Fig:OpticalConductivity}. We observe that, for all the compounds there is a threshold for the onset of optical transitions that is equal to the gap $\Delta$. The steplike structure of $\sigma(\omega)$ at low energies (see insets of each panel in Fig. \ref{Fig:OpticalConductivity}) is due to the SOC, that leads to two set of optical transitions in the spectrum. Due to the stronger SOC of heavier W atoms, the effect is specially visible in WS$_2$ and WSe$_2$, with plateaus of $\sim 0.4$~eV in $\sigma(\omega)$, corresponding to the energy splitting of the valence band at the K and K' points of the Brillouin zone (see Fig. \ref{Fig:Fitting}). These transitions lead to the well known A and B absorption peaks observed in photoluminescence.\cite{MH12} We further notice that our results for the optical conductivity are in good agreement, even for the onset energy, with experimental measurements (see e.g. Ref. \onlinecite{MH10}). At higher energies, the optical conductivity shows a series of peaks that are associated to optical transitions between flat bands in the spectrum (van Hove singularities). Such van Hove singularities are clearly evident for the valence and conduction bands at the M point of the Brillouin zone (see Fig. \ref{Fig:Fitting}). We notice that disorder (vacancies, adatoms, etc.), not included here, can lead to the creation of midgap states that allows for additional optical transitions in the spectrum.\cite{YG14}

\section{Summary}

In summary, we have generalized the tight-binding model for MoS$_2$ of Ref. \onlinecite{CG13} to the other families of semiconducting TMDs: MoSe$_2$, WS$_2$ and WSe$_2$. Our main result is the set of Slater-Koster parameters of Table \ref{Tab:Parameters}, which have been obtained from a fit to DFT calculations in which special care was paid to capture the main orbital contribution of the TB bands at the relevant regions of the band structure.  The obtained models have been used to calculate the optical conductivity of the different compounds. This approximation can be straightforwardly  generalized to multi-layer systems with arbitrary stacking orders, heterostructures made from the stacking of layers of different compounds, twisted multilayers, strained and/or disordered samples, etc.

\acknowledgments

We appreciate useful conversations with E. Cappelluti, P. Ordej\'on, F. Guinea, M.P. L\'opez-Sancho and H. Rostami.  J.A. S.-G. acknowledges financial support from European Union's Seventh Framework Programme (FP7/2007- 2013) through the ERC Advanced Grant NOV- GRAPHENE (GA 290846).  P. S-J. was supported by MINECO through Grant No. FIS2015-65706-P and the Ram\'on y Cajal programme RYC-2013-14645. R.R. acknowledges financial support from MINECO (FIS2014-58445-JIN).

\bibliography{biblio.bib}

\appendix

\section{On-site and hopping matrices of the $6\times 6$ block}\label{App:TB}

For convenience we reproduce in this appendix the analytical expressions for the model. The on-site terms of the Hamiltonian can be written in the compact form:\cite{Roldan_2DMat_2014}
\begin{eqnarray}
\boldsymbol{\epsilon}
&=&
\left(
\begin{array}{cc}
\epsilon_M & 0 \\
0 & \epsilon_X
\end{array}
\right),
\end{eqnarray}
where
\begin{eqnarray}
\epsilon_M
&=&\begin{pmatrix}\Delta_0&&0&&0\\0&&\Delta_2&&-i\lambda_M\hat{s}_z\\0&&i\lambda_M\hat{s}_z&&\Delta_2\end{pmatrix},
\nonumber\\
\nonumber\\
\epsilon_X
&=&\begin{pmatrix}\Delta_p+t^\perp_{xx}&&-i\frac{\lambda_X}{2}\hat{s}_z&&0\\i\frac{\lambda_X}{2}\hat{s}_z&&\Delta_p+t^\perp_{yy}&&0\\0&&0&&\Delta_z-t^\perp_{zz}\end{pmatrix},
\end{eqnarray}
where $\lambda_M$ and $\lambda_X$ are the SOC of the metal ($M$) and chalcogen atoms ($X$), respectively, and $\hat{s}_z=\pm$ is the $z$-component of the spin degree of freedom.\cite{Roldan_2DMat_2014}
The effects of vertical hopping $V_{pp}$
between top and bottom $X$ atoms is considered through the terms $t^\perp_{xx}=t^\perp_{yy}=V_{pp\pi}$, and $t^\perp_{zz}=V_{pp\sigma}$. The nearest neighbor hopping between $M$ and $X$ atoms are
\begin{widetext}
\begin{align}
t^{MX}_1&=\frac{\sqrt{2}}{7\sqrt{7}}\begin{pmatrix}-9 V_{pd\pi}+\sqrt{3}V_{pd\sigma}&&3\sqrt{3}V_{pd\pi}-V_{pd\sigma}&&12 V_{pd\pi}+\sqrt{3} V_{pd\sigma}\\
          5\sqrt{3} V_{pd\pi}+3 V_{pd\sigma}&&9 V_{pd\pi}-\sqrt{3} V_{pd\sigma}&&-2\sqrt{3}V_{pd\pi}+3 V_{pd\sigma}\\
          -V_{pd\pi}-3\sqrt{3}V_{pd\sigma}&&5\sqrt{3}V_{pd\pi}+3 V_{pd\sigma}&&6 V_{pd\pi}-3\sqrt{3} V_{pd\sigma}\end{pmatrix}\\
\nonumber\\
t^{MX}_2&=\frac{\sqrt{2}}{7\sqrt{7}}\begin{pmatrix}0&&-6\sqrt{3}V_{pd\pi}+2V_{pd\sigma}&&12V_{pd\pi}+\sqrt{3}V_{pd\sigma}\\
          0&&-6V_{pd\pi}-4\sqrt{3}V_{pd\sigma}&&4\sqrt{3} V_{pd\pi}-6V_{pd\sigma}\\14V_{pd\pi}&&0&&0\end{pmatrix}\\
\nonumber\\
t^{MX}_3&=\frac{\sqrt{2}}{7\sqrt{7}}\begin{pmatrix}9 V_{pd\pi}-\sqrt{3}V_{pd\sigma}&&3\sqrt{3}V_{pd\pi}-V_{pd\sigma}&&12 V_{pd\pi}+\sqrt{3} V_{pd\sigma}\\
          -5\sqrt{3} V_{pd\pi}-3 V_{pd\sigma}&&9 V_{pd\pi}-\sqrt{3} V_{pd\sigma}&&-2\sqrt{3}V_{pd\pi}+3 V_{pd\sigma}\\
          -V_{pd\pi}-3\sqrt{3}V_{pd\sigma}&&-5\sqrt{3}V_{pd\pi}-3 V_{pd\sigma}&&-6V_{pd\pi}+3\sqrt{3} V_{pd\sigma}\end{pmatrix}
\end{align}

\end{widetext}
where the direction of the hopping labelled by subindices 1,2, and 3 is shown in Fig.~\ref{Fig:Structure}(b). Hopping terms corresponding to processes between the same kind of atoms, $M$-$M$ or $X$-$X$ (see Fig. \ref{Fig:Structure}(c)), are given by
\begin{widetext}
\begin{align}
t^{MM}_1&=\frac{1}{4}\begin{pmatrix}3V_{dd\delta}+V_{dd\sigma}&&\frac{\sqrt{3}}{2}(-V_{dd\delta}+V_{dd\sigma})&&-\frac{3}{2}(V_{dd\delta}-V_{dd\sigma})\\
        \frac{\sqrt{3}}{2}(-V_{dd\delta}+V_{dd\sigma})&&\frac{1}{4}(V_{dd\delta}+12V_{dd\pi}+3V_{dd\sigma})&&\frac{\sqrt{3}}{4}(V_{dd\delta}-4V_{dd\pi}+3V_{dd\sigma})\\
        -\frac{3}{2}(V_{dd\delta}-V_{dd\sigma})&&\frac{\sqrt{3}}{4}(V_{dd\delta}-4V_{dd\pi}+3V_{dd\sigma})&&\frac{1}{4}(3V_{dd\delta}+4V_{dd\pi}+9V_{dd\sigma})\end{pmatrix}\\
\nonumber\\
t^{MM}_2&=\frac{1}{4}\begin{pmatrix}3V_{dd\delta}+V_{dd\sigma}&&\sqrt{3}(V_{dd\delta}-V_{dd\sigma})&&0\\\sqrt{3}(V_{dd\delta}-V_{dd\sigma})&&V_{dd\delta}+3V_{dd\sigma}&&0\\0&&0&&4V_{dd\pi}\end{pmatrix}\\
\nonumber\\
t^{MM}_3&=\frac{1}{4}\begin{pmatrix}3V_{dd\delta}+V_{dd\sigma}&&\frac{\sqrt{3}}{2}(-V_{dd\delta}+V_{dd\sigma})&&\frac{3}{2}(V_{dd\delta}-V_{dd\sigma})\\
        \frac{\sqrt{3}}{2}(-V_{dd\delta}+V_{dd\sigma})&&\frac{1}{4}(V_{dd\delta}+12V_{dd\pi}+3V_{dd\sigma})&&-\frac{\sqrt{3}}{4}(V_{dd\delta}-4V_{dd\pi}+3V_{dd\sigma})\\
        \frac{3}{2}(V_{dd\delta}-V_{dd\sigma})&&-\frac{\sqrt{3}}{4}(V_{dd\delta}-4V_{dd\pi}+3V_{dd\sigma})&&\frac{1}{4}(3V_{dd\delta}+4V_{dd\pi}+9V_{dd\sigma})\end{pmatrix}\\
\nonumber\\
t^{XX}_1&=\frac{1}{4}\begin{pmatrix}3V_{pp\pi}+V_{pp\sigma}&&\sqrt{3}(V_{pp\pi}-V_{pp\sigma})&&0\\
                                     \sqrt{3}(V_{pp\pi}-V_{pp\sigma})&&V_{pp\pi}+3V_{pp\sigma}&&
                                     0\\0&&0&&4V_{pp\pi}\end{pmatrix}\\
\nonumber\\
t^{XX}_2&=\begin{pmatrix}V_{pp\sigma}&&0&&0\\0&&V_{pp\pi}&&0\\0&&0&&V_{pp\pi}\end{pmatrix}\\
\nonumber\\
t^{XX}_3&=\frac{1}{4}\begin{pmatrix}3V_{pp\pi}+V_{pp\sigma}&&-\sqrt{3}(V_{pp\pi}-V_{pp\sigma})&&0\\-\sqrt{3}(V_{pp\pi}-V_{pp\sigma})&&V_{pp\pi}+3V_{pp\sigma}
                                      &&0\\0&&0&&4V_{pp\pi}\end{pmatrix}.
\end{align}
\end{widetext}

\section{Orbital contribution of the tight-binding bands}

\begin{figure*}
\includegraphics[scale=0.28,clip=]{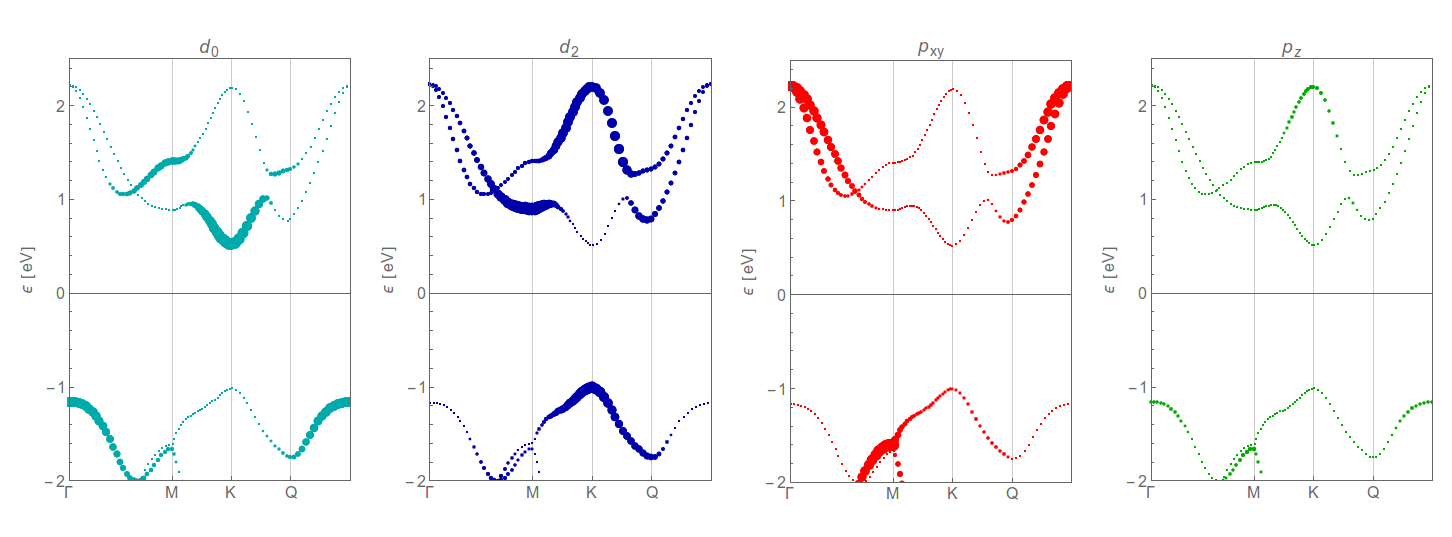}
\includegraphics[scale=0.28,clip=]{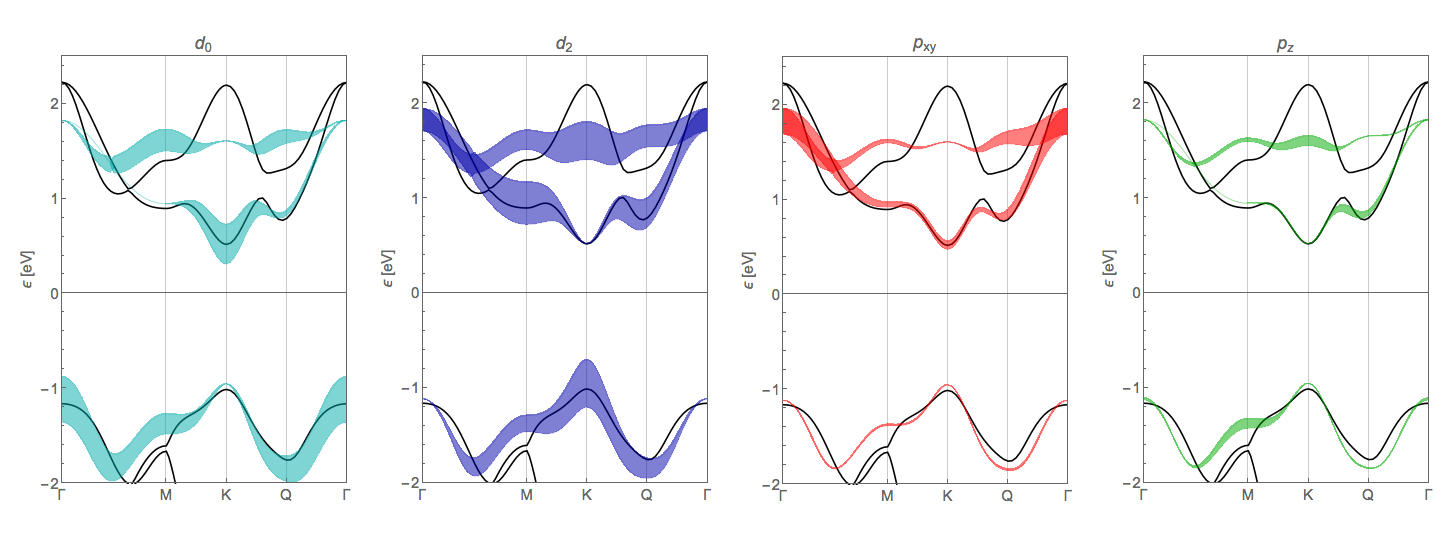}
\caption{Same as Fig. \ref{Fig:FB-MoS2} but for MoSe$_2$. 
}
\label{Fig:FB-MoSe2}
\end{figure*}

\begin{figure*}[t]
\includegraphics[scale=0.28,clip=]{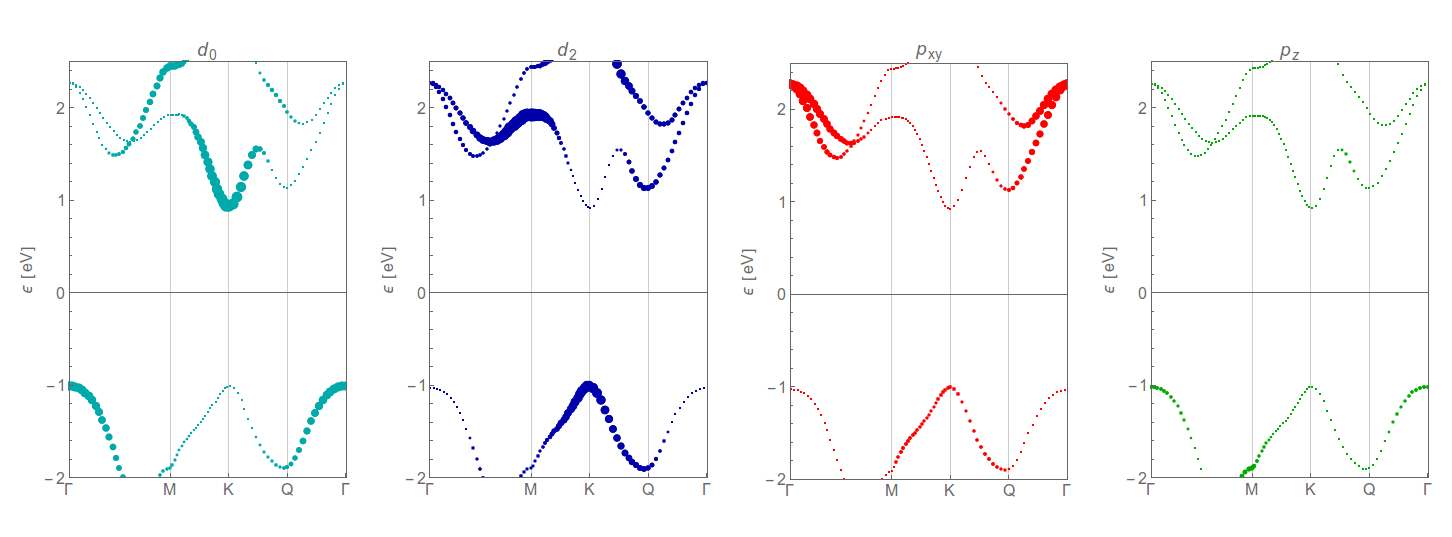}
\includegraphics[scale=0.28,clip=]{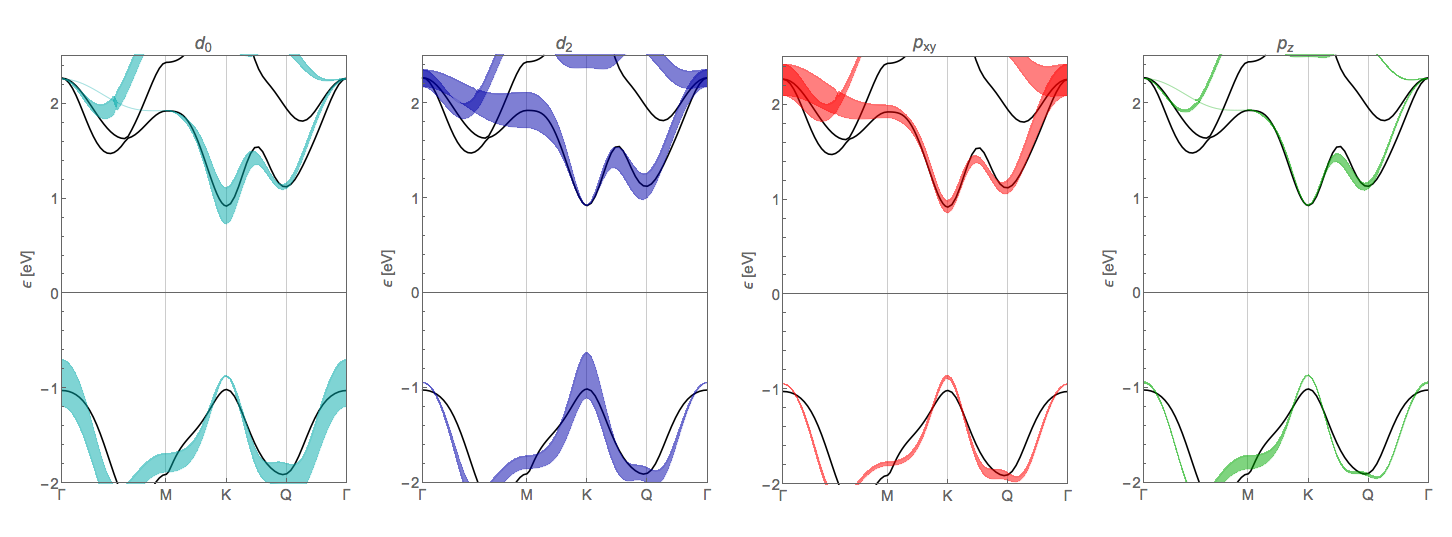}
\caption{Same as Fig. \ref{Fig:FB-MoS2} but for WS$_2$. 
}
\label{Fig:FB-WS2}
\end{figure*}

\begin{figure*}[t]
\includegraphics[scale=0.28,clip=]{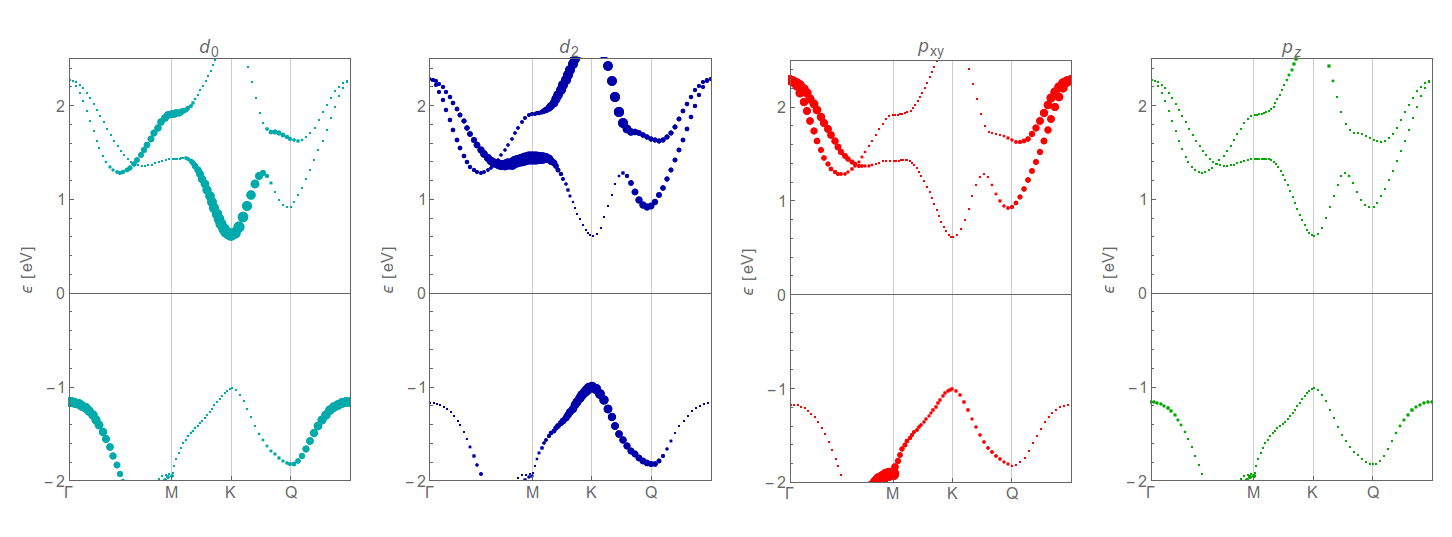}
\includegraphics[scale=0.28,clip=]{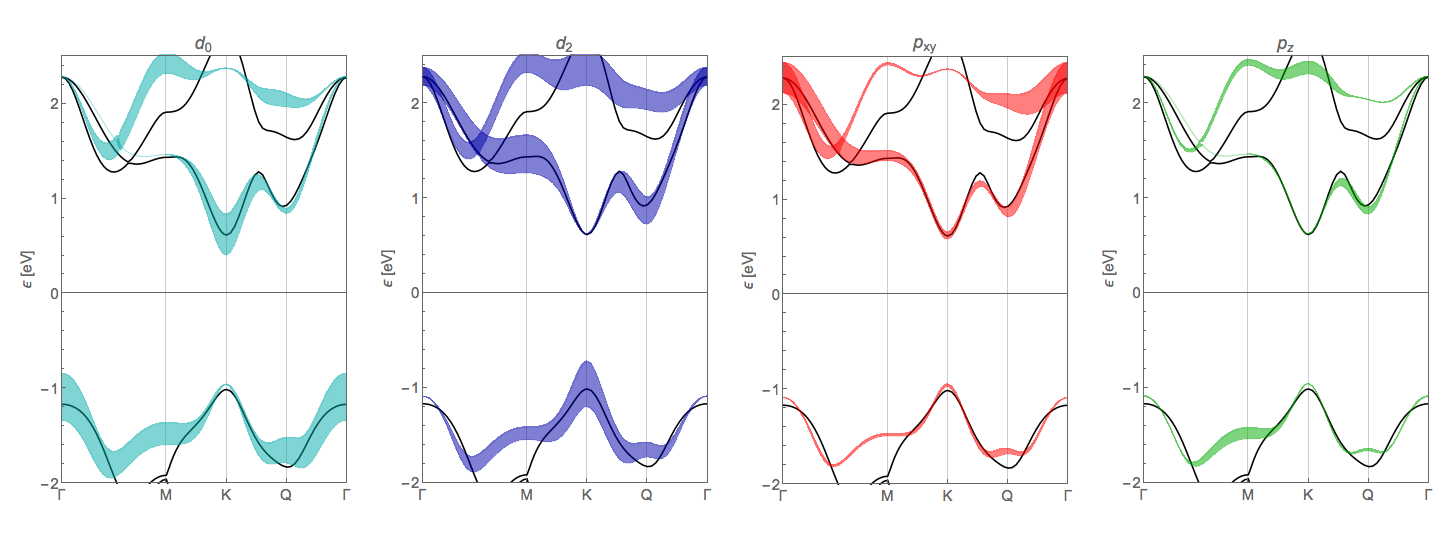}
\caption{Same as Fig. \ref{Fig:FB-MoS2} but for WSe$_2$. 
}
\label{Fig:FB-WSe2}
\end{figure*}

In this appendix we show the orbital contribution of the tight-binding bands, as compared to the DFT results, for MoSe$_2$ (Fig. \ref{Fig:FB-MoSe2}), WS$_2$ (Fig. \ref{Fig:FB-WS2}) and WSe$_2$ (Fig. \ref{Fig:FB-WSe2}). 

\end{document}